# Resilient Operational Planning for Microgrids Against Extreme Events

Cunzhi Zhao
University of Houston
czhao20@uh.edu

Jesus Silva-Rodriguez
University of Houston
jasilvarodriguez@uh.edu

Xingpeng Li
University of Houston
xli82@uh.edu

## Abstract

*This paper proposes a novel resilience index, a microgrid survivability rate (SR) under extreme events, and then proposes a novel Resilient Operational Planning (ROP) algorithm to maximize the proposed resilience index SR. The proposed ROP algorithm can incorporate predetermined inverter failure probabilities and generate multiple scenarios accordingly to optimize resilient operations during an extreme event. The implemented ROP algorithm consists of two main steps: (i) optimization of resilient operational planning, and (ii) preventive resilience enhancement if minimum SR is not met per the analysis in step 1. A typical microgrid (MG) is studied to compare the proposed ROP algorithm against a traditional microgrid energy management (MEM) model. Results indicate that an enhanced resilience operation is achieved by the ROP algorithm, which is demonstrated by the quantification of resilience via the SR. Moreover, the proposed ROP algorithm is able to obtain a greater SR overall compared to that achieved by the traditional MEM, and this benefit of using the proposed ROP increases as the inverter failure probabilities increase.*

## 1. Introduction

There has been an increase in extreme events in the last few years that negatively and substantially impact our power system infrastructures, affecting generation, transmission, and distribution of electric power, leading to disruptions of the power services to consumers. Consequently, it is important that future power system models consider methods for resilience enhancement.

Microgrid is a low-voltage local power system consisting of controllable distributed energy resources (DERs) and renewable DERs over a small territory [1-4]. Therefore, a microgrid on its own is inherently a resilience enhancement resource for the main power grid thanks to its ability to isolate and meet its local demand independently when an extreme event affects the transmission network or other important infrastructure of the main power systems [5]. However, microgrids are also vulnerable to these extreme events and can present issues internally that need to be addressed by appropriate resilience enhancement methods. Under extreme weather scenarios, there is an increased inverter failure probability that can largely affect the microgrid resilience. Therefore, it is important to consider and design operational models that include inverter failure probabilities to accurately determine the total generation that will be available during an extreme event.

A state-of-the-art overview showed that standard power system reliability assessments exclude the wear-out failure of power electronics and concluded that for reliability-oriented design and planning for microgrids involving a large installed capacity of power electronics-interfaced DERs, it is crucial to consider their probability of failure [6]. Still, even though there is substantial research in the area of microgrid system resilience and strategies to ensure microgrids can continue to serve their most critical loads during extreme events, most of the research focuses on strategies related to prior scheduling of resources, system reconfigurations, component-based outage management, and forecasting methods to predict severity and level of disruption during these events [5]. Very limited research efforts investigate microgrid management considering the failure of inverter-based DERs and how to implement these to achieve a more resilient operational planning.

Holistic approaches for assessing threats and vulnerabilities in a microgrid as well as defining resilience metrics are proposed in [7] and [8]. The resilience metrics are defined based on a quantification of microgrid properties associated to the impact of an extreme event, and are defined as threats, vulnerability, and vulnerability impacts, each dependent on the other. These quantified properties are used to determine a risk parameter that can be applied to determine a resilience metric. However, both [7] and [8] do not consider probabilities of inverter failures or other components associated to the power generation of the microgrid.

A statistical framework to quantify the resilience of a microgrid and ensure critical demand is met during islanding scenarios is developed in [9], where the system-level resilience metric uses asset-level reliability data and develops a dispatch routine for islanded operation to improve microgrid resilience. However, [9] focuses on optimization of available assets and implements dispatching strategies subject to failure rates in real-time once the disruption event is in progress. There is no consideration of inverter failure-induced loss of DER, nor does it prepare the microgrid for possible scenarios beforehand.

A resilience-constrained operation strategy is proposed in [10] which uses a battery energy storage system (BESS) as a resilience resource by defining resiliency cuts for the state of charge of the BESS units. Although these resiliency cuts ensure the survivability of critical load by setting charge level limits in the BESS, this approach does not consider the probability of failure of the BESS themselves or the inverters associated with those units.

A holistic operation failure rate model of power electronics systems based on overall reliability assessment is performed in [11], which features a component reliability model that involves empirical and physical models to determine operation failure rate of devices, which are then used as reliability metrics applied to a short-term outage model. Although [11] does consider failure of the power electronics of the DER, it mainly focuses on the system reliability, and it does not consider system resilience for extreme events outside a short-term low-impact outage.

Since inverter failure can lead to the loss of interconnected energy sources and hence affect a microgrid's reliability and power supply capability, this paper proposes a Resilient Operational Planning (ROP) algorithm that optimizes a resilience metric for microgrid survivability (SR). The ROP algorithm incorporates predicted values of inverter failure probability at different stages of an extreme event to minimize the loss of power supply to critical loads and maximize the microgrid's resilience.

The remainder of the paper is organized as follows. Section 2 presents and describes the ROP optimization model and how it improves the resilient operation of a microgrid based on inverter failure information. Section 3 describes the functionality of the ROP's algorithm, how it generates the scenarios, and how it enhances resilience. Section 4 presents a case study composed of multiple scenarios based on the number of inverters and their failure probabilities and compares the microgrid's resilience metric SR between the proposed ROP model and a regular microgrid energy management (MEM) model to showcase how the ROP is able to provide a reduced loss of critical load during an extreme event. Finally, section 5 concludes the paper.

## 2. Mathematical Formulation

The ROP algorithm presented in this paper supports decision-making for three stages of an extreme event: pre-contingency, during-contingency, and post-contingency phases. A defined time period $T$ of the extreme event is divided into three sub-sets for the duration of these three stages, and inverters are assigned different probabilities of failure for each stage, with the during-contingency phase having the highest probability of failure. Based on these failure probabilities at each time interval $t$, a large number of scenarios is generated consisting of instances at which, if the particular scenario occurs, the inverter will fail at the end of that time interval if used.

### 2.1. Sets and Indices

$t$: Time period index.
$T$: Time period set.
$g$: Controllable DER index.
$G$: Controllable DER set.
$\omega$: Scenario index.
$\Omega$: Scenario set.
$i$: Inverter index.
$I$: Inverter set.

### 2.2. Variables

$v^{I}_{i,t,\omega}$: On/off status indicator for inverter $i$.
$P^{G}_{g,t,\omega}$: Power output of generator $g$.
$u^{G}_{g,t}$: On/off status indicator for generator $g$.
$u^{I}_{i,t,\omega}$: Availability indicator for inverter $i$.
$X^{T}_{t,\omega}$: Successful indicator of time interval $t$ of scenario $\omega$.
$X_{\omega}$: Successful indicator of scenario $\omega$.
$Y_{t,\omega}$: Surplus variable.
$P^{NET}_{t,\omega}$: Aggregated power of scenario $\omega$ at time interval $t$.
$w^{I}_{i,t,\omega}$: Confirmed inverter failure indicator for inverter $i$.

### 2.3. Parameters

$\rho_{\omega}$: Probability of scenario $\omega$.

$P^G_{min,g}$: Minimum power output of generator *g*.
$P^G_{max,g}$: Maximum power output of generator *g*.
$P^I_{i,t}$: Power output from inverter *i* (solar/wind power).
$f^I_{i,t,\omega}$: Failure indicator for inverter *i*.
$P^L_{crit,t}$: Critical power load.
$R^G_g$: Ramping limit of generator g.
$\alpha$: Minimum acceptable percentage of critical load supplied under extreme events.

## 2.4. Optimization Model

The objective of the ROP optimization model is to maximize the SR resilience metric by maximizing the number of scenarios where the power supplied by the microgrid never drops below a predefined percentage of critical load throughout the time period *T*. This is mathematically described by (1).

$$maximize\ F = \sum_{\omega \in \Omega} X_\omega \rho_\omega \quad (1)$$

As a result of the consideration of an extreme event such as a high category hurricane or a severe winter storm, the microgrid will be considered to be in islanded operation, assuming the main grid failed as a result of the disaster. The microgrid will only rely on controllable DERs, such as diesel generators (DG), micro-turbines (MT), and fuel cells (FC), and a set of inverters associated to renewable DERs to meet its demand. The inverters are subject to fail and the output of the optimization model will determine the decision of which distributed generators to dispatch to cover all possible scenarios, as well as which inverters to keep online during the time intervals of the disruptive event to ensure a high reliability in as many scenarios as possible and explore how the acceptable percentage threshold $\alpha$ of critical load affects the number of possible successful scenarios.

Because the microgrid will feature controllable generators, constraints (2)-(3) represent the output limits and ramping limits of each generator, respectively. Constraints (4)-(9) model the availability, failure, and decision of utilizing the inverters at each time interval for every scenario. When the inverter is destined to fail and it is used, that inverter becomes unavailable for the next time interval and remains unavailable for the rest of the time period, and if it is not used, it will remain available. Equation (10) defines the total available power in the microgrid to supply the critical load, and (11)-(13) determine whether the critical load was successfully met at that particular time interval *t* for that scenario *ω*, implementing the "*Big-M*" strategy on constraint (11). Finally, (14)-(15) define whether a scenario can be considered successful: if it does not present any unsuccessful time interval when the available power drops below the user-prespecified acceptable threshold of critical load.

$$P^G_{min,g} u^G_{g,t} \le P^G_{g,t,\omega} \le P^G_{max,g} u^G_{g,t}\ ,\ \forall g \in G, t \in T, \omega \in \Omega \quad (2)$$

$$-R^G_g \le P^G_{g,t,\omega} - P^G_{g,t-1,\omega} \le R^G_g\ ,\ \forall g \in G, t \in T, \omega \in \Omega \quad (3)$$

$$v^I_{i,t,\omega} \le u^I_{i,t,\omega}\ ,\ \forall i \in I, t \in T, \omega \in \Omega \quad (4)$$

$$u^I_{i,t,\omega} \le u^I_{i,t-1,\omega}\ ,\ \forall i \in I, t \in T, \omega \in \Omega \quad (5)$$

$$w^I_{i,t,\omega} \ge v^I_{i,t,\omega} + f^I_{i,t,\omega} - 1\ ,\ \forall i \in I, t \in T, \omega \in \Omega \quad (6)$$

$$w^I_{i,t,\omega} \le v^I_{i,t,\omega}\ ,\ \forall i \in I, t \in T, \omega \in \Omega \quad (7)$$

$$w^I_{i,t,\omega} \le f^I_{i,t,\omega}\ ,\ \forall i \in I, t \in T, \omega \in \Omega \quad (8)$$

$$u^I_{i,t+1,\omega} \le 1 - w^I_{i,t,\omega}\ ,\ \forall i \in I, t \in T, \omega \in \Omega \quad (9)$$

$$\sum_{g \in G} P^G_{g,t,\omega} + \sum_{i \in I} P^I_{i,t} v^I_{i,t,\omega} = P^{NET}_{t,\omega}\ ,\ \forall t \in T, \omega \in \Omega \quad (10)$$

$$Y_{t,\omega} \le P^{NET}_{t,\omega} - \alpha P^L_{crit,t} + M \cdot \left(1 - X^T_{t,\omega}\right)\ ,\ \forall t \in T, \omega \in \Omega \quad (11)$$

$$Y_{t,\omega} \ge P^{NET}_{t,\omega} - \alpha P^L_{crit,t}\ ,\ \forall t \in T, \omega \in \Omega \quad (12)$$

$$0 \le Y_{t,\omega} \le M X^T_{t,\omega}\ ,\ \forall t \in T, \omega \in \Omega \quad (13)$$

$$X_\omega \le X^T_{t,\omega}\ ,\ \forall t \in T, \omega \in \Omega \quad (14)$$

$$\sum_{t \in T} X^T_{t,\omega} - (T-1) \le X_\omega\ ,\ \forall \omega \in \Omega \quad (15)$$

Since there are probabilities and scenarios involved, this model becomes a stochastic optimization problem, with the probability of each scenario determined based on the combination of the individual failure probabilities at each time interval for each inverter.

To set a case for comparison, the loss of critical load of the ROP will be compared with the critical load loss resulting from the implementation of a regular MEM [12-14].

# 3. Resilient Operational Planning Algorithm

There are two main steps of the proposed ROP algorithm, resilient operational planning optimization and SR enhancement. The ROP algorithm flowchart is presented in Figure 1. In the ROP optimization step, the scenarios are generated based on the given inverter failure probabilities following a scenario generation algorithm that is illustrated in Algorithm 1.

The generated scenarios involving the inverter failure information are used as input to the ROP optimization model, and probability is determined for each scenario. $X_\omega$ would be 1 only when the aggregated power is enough to supply the critical load for all time intervals in scenario $\omega$. The ROP optimization of section 2 will maximize the expectation of $X_\omega$ for all the input scenarios. The resilience metric SR refers to the expectation value of $X_\omega$, which represents the possibility that the MG can supply the critical load under the given inverter failure probabilities.

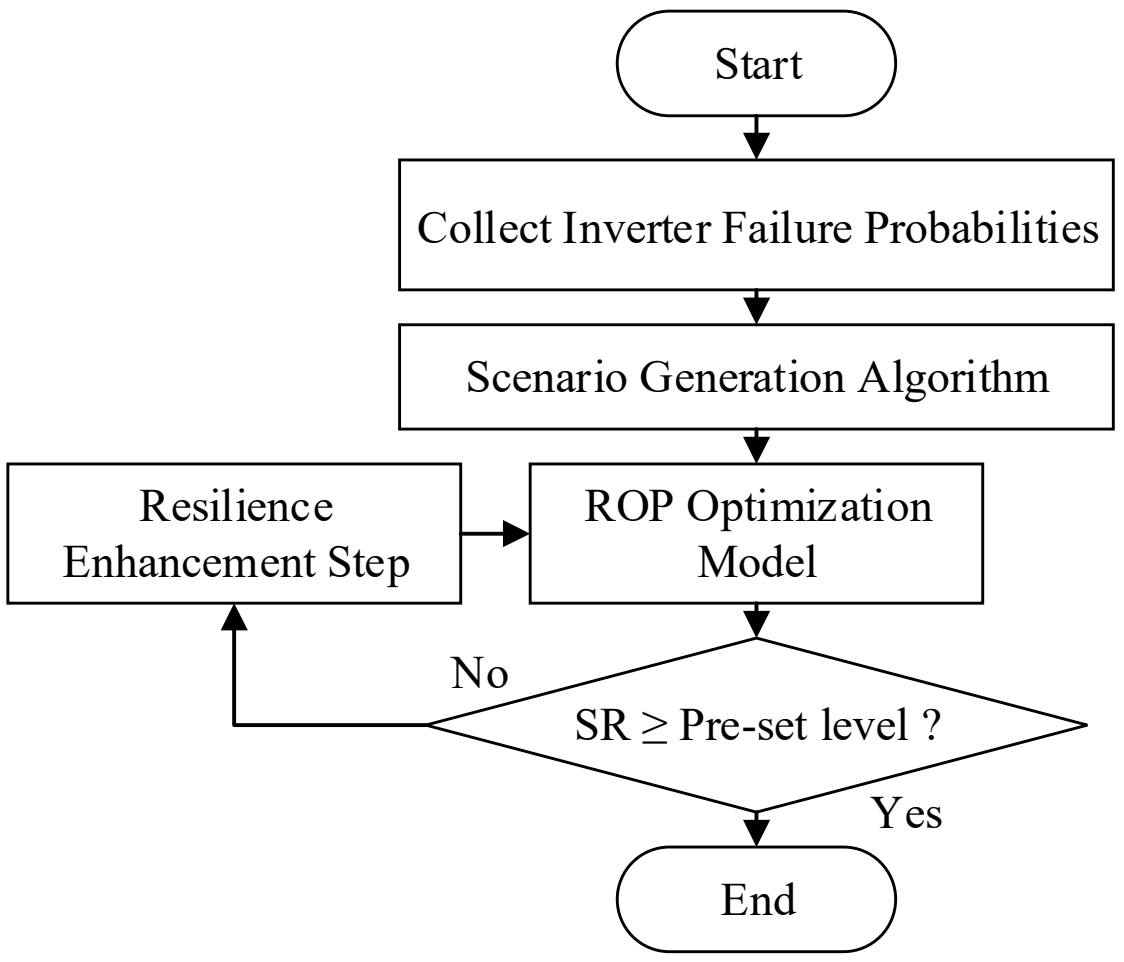

**Figure 1. ROP algorithm flowchart.**

**Algorithm 1: Scenario generation of ROP algorithm.**

1. Set the total time interval $T$ and number of inverters $I$.
2. Obtain the failure probability for each inverter.
3. Set the desired scenario number $\Omega$ and initialize $\omega = 1$.
4. **If** $\omega \le \Omega$, create a matrix of $(T,I)$.
5. **Else** Stop Algorithm.
6. **For** $t$ *in* $T$
7. **For** $i$ *in* $I$
8. Set $X = random(0,1)$
9. **If** $X \le inverter\ failure\ probability$, set $(t,i) = 1$.
10. **Else** Set $(t,i) = 0$.
11. **end For**
12. **end For**
13. $\omega = \omega + 1$, and go to step 4.

The SR enhancement step is activated when the value of SR from the first step does not meet the pre-set resilience level. For instance, if the SR value obtained from the first step is 90% which is less than the pre-set 95%, the SR enhancement step will dispatch an extra portable DG to the MG system, so the updated ROP optimization can obtain a higher value of SR. If the updated SR meets the requirement, the algorithm stops, and if not, the SR enhancement step will be activated again.

## 4. Case Studies

To validate the effectiveness of the proposed ROP algorithm, a typical islanded MG system is modeled in this paper. The testbed MG system includes three controllable DER units which are MT, FC, and DG. The parameters of the controllable DERs are shown in Table 1. Ten inverter-based renewable DER units including wind turbine and solar photovoltaics are modeled in the MG system. Only the inverter-based renewable DERs are assumed to be subject to fail in this case study. The load data which represents a community microgrid with 1000 residential houses is collected by Pecan Street Dataport [15]. The critical load is set to 20% of the total load. The failure percentage of inverter is assumed to be provided by a third party. The scenarios of inverter failure are generated using the provided failure probabilities. The ROP algorithm is implemented and solved in Python with the Pyomo package [16] and Gurobi Optimizer Solver [17].

**Table 1. Controllable DER Parameters.**

| DER | $P^G_{min,g}$ | $P^G_{max,g}$ | $R^G_g$ | Op-Cost | NL Cost | SU Cost |
|---|---|---|---|---|---|---|
| MT | 18 kW | 180 kW | 240 kW/h | 0.08 \$/kWh | \$3.4 | \$5 |
| FC | 12.7 kW | 75 kW | 280 kW/h | 0.18 \$/kWh | \$1.74 | \$3.5 |
| DG | 14 kW | 80 kW | 170 kW/h | 0.16 \$/kWh | \$2 | \$5 |

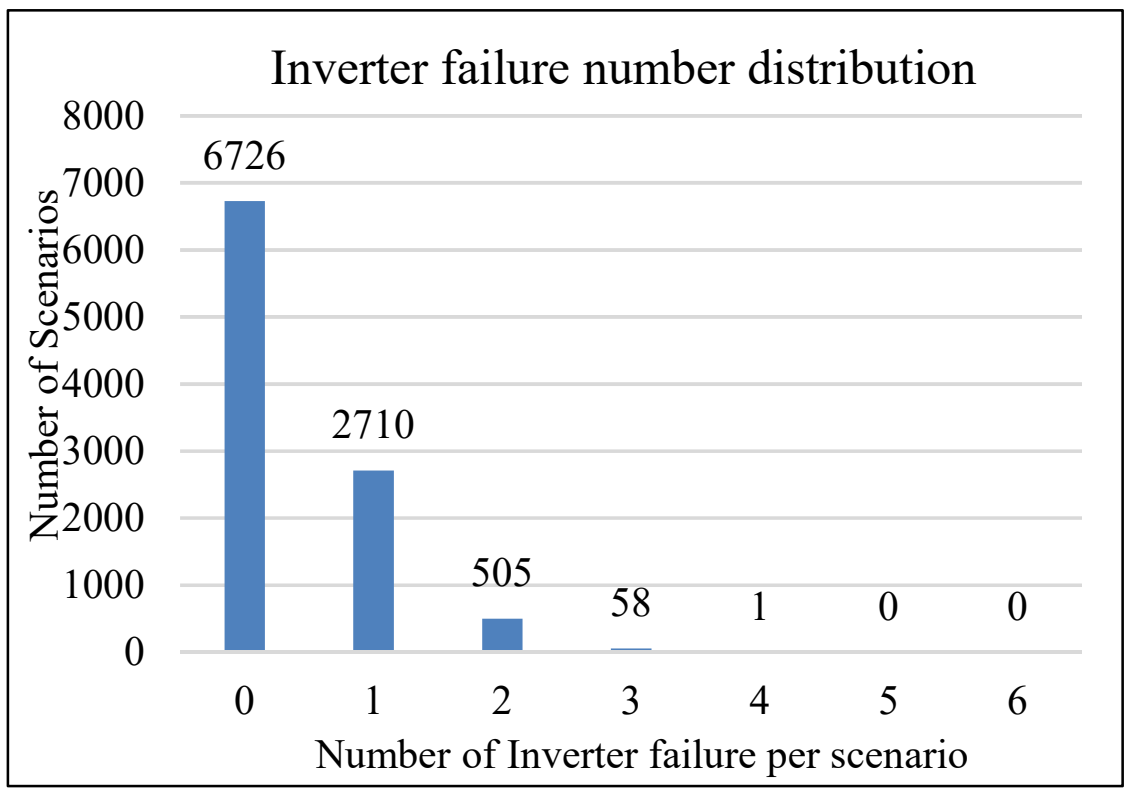

**Figure 2. Sample of inverter failure number distribution.**

Figure 2 presents a distribution of the total number of inverter failures under certain failure probability. A total of 10,000 scenarios is generated, and all the inverter failure probabilities are assumed to be 1% in this example. From Figure 2, it can be observed that most generated scenarios have zero inverter failures. Also, there is no scenario that contains 5 or more inverter failures at the same time.

All the sensitivity tests and results shown in the rest of the paper are conducted with generated scenarios like those shown in Figure 2. The scenarios for each test differ by the inverter failure probabilities and time intervals.

Table 2 presents 6 cases of inverter failures. From C1 to C6, the inverter failure probabilities increase gradually. Each case's probabilities are applied to generate a group of scenarios to evaluate the performance of the proposed algorithm. The length of the time interval is four hours. Besides the ROP algorithm, a benchmark MEM model is also set, whose objective is to minimize the total operating cost. The term SR is used to refer to the expectation value of the probability that the MG system can support enough power to a percentage of critical load. The percentage of critical load $\alpha$ is set to 100% in all the tests in Table 2. To evaluate the performance of the proposed optimization model, all values of SR are based on the ROP algorithm that excludes the resilient enhancement step in Table 1. The SR of C1 is 95% and is expected to decline steadily with the gradual increase of inverter failure probabilities from C1 to C6. The results shown in Table 2 meet the expectation, and these cases will be used to model the different events for further analysis of the proposed ROP algorithm.

**Table 2. Different cases of inverter failure and related probability.**

| | Case | C1 [%] | C2 [%] | C3 [%] | C4 [%] | C5 [%] | C6 [%] |
|---|---|---|---|---|---|---|---|
| Prob. of inverter failure (10 Inverters) | 1 | 0.5 | 1 | 1 | 2 | 3 | 5 |
| | 2 | 0.5 | 1 | 1 | 2 | 3 | 5 |
| | 3 | 0.5 | 1 | 1 | 2 | 3 | 5 |
| | 4 | 0.5 | 1 | 1 | 2 | 3 | 5 |
| | 5 | 0.5 | 1 | 1 | 2 | 3 | 5 |
| | 6 | 0.5 | 1 | 2 | 2 | 3 | 5 |
| | 7 | 0.5 | 1 | 2 | 2 | 3 | 5 |
| | 8 | 0.5 | 1 | 2 | 2 | 3 | 5 |
| | 9 | 0.5 | 1 | 2 | 2 | 3 | 5 |
| | 10 | 0.5 | 1 | 2 | 2 | 3 | 5 |
| Microgrid SR | ROP | 95 | 90 | 86 | 82 | 73 | 58 |
| | MEM | 86 | 73 | 63 | 55 | 40 | 22 |

Figure 3 compares the summary statistics of SR for ROP and MEM according to Table 2. As can be seen from the figure, both groups of SR decrease from C1 to C6. However, for the same cases, the ROP algorithm can achieve substantially higher SR than the MEM algorithm. Also, the difference in SR between ROP and MEM increases from C1 to C6. In other words, when the inverter failure probabilities increase, the proposed ROP has greater ability to enhance the value of SR compared to the MEM algorithm. From Figure 3, we can find that the performance of the ROP algorithm is solid and can improve the reliability of an MG significantly.

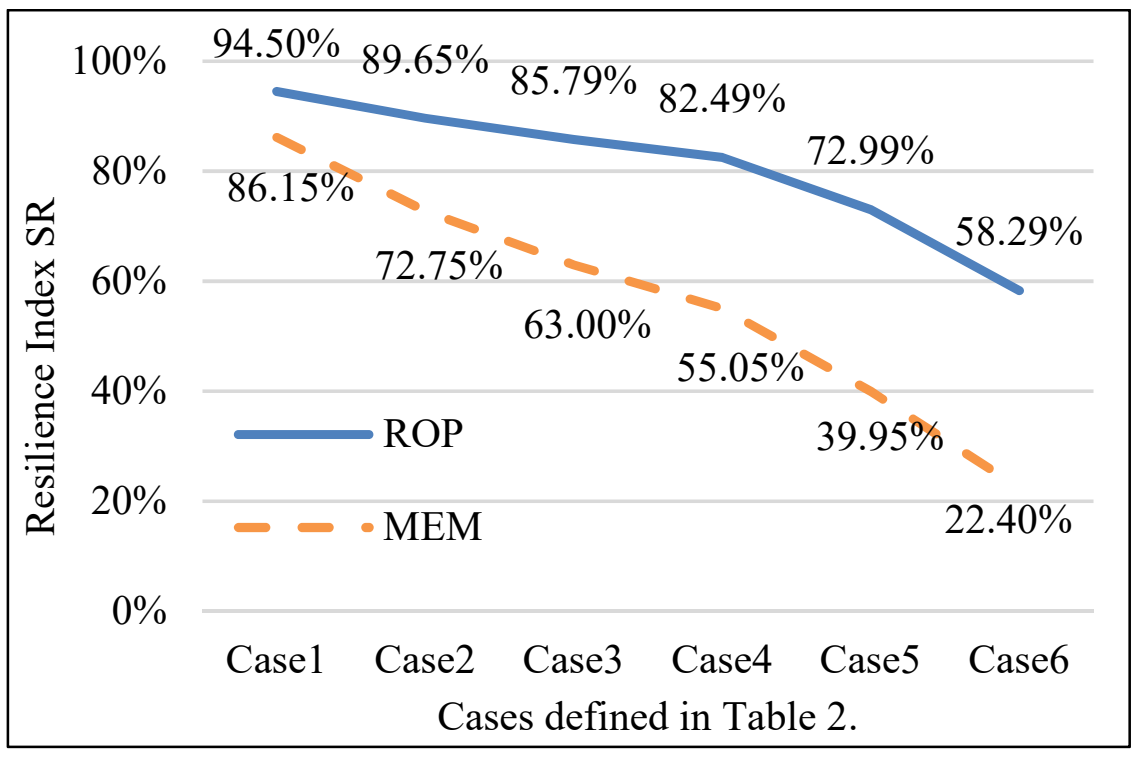

**Figure 3. Plot of survivability rate.**

Table 3 presents the correlation between the value of SR and the pre-set $\alpha$ value. The probabilities of inverter failure are set to those of C1 from Table 2. The length of time interval is set to four hours. On the other hand, the value of SR will always be 100% when $\alpha$ value is below 95%. When $\alpha$ is 98%, the value of SR barely decreases to 99.5%, which means the MG is able to supply 98% power of critical load for almost all the generated scenarios. When $\alpha$ increases to 100%, the value of SR decreases to 94.5%. If we further increase $\alpha$ to 105% the MG fails to supply enough power due to the aggregated available power being less than 105% of the critical load even when there is no inverter failure.

**Table 3. Survivability Rate sensitivity test of $\alpha$.**

| $\alpha$ | 90% | 95% | 98% | 100% | 102% | 105% |
|---|---|---|---|---|---|---|
| SR | 100% | 100% | 99.5% | 94.5% | 94.2% | 0% |

To further evaluate the ROP algorithm, different events are created with different inverter failure probabilities and longer time intervals. There are a total of three events as shown in Figure 4. All events are designed with a 12-hour time period. The 12 hours are split into three stages: pre-contingency (hour 1-4); on-contingency (hour 5-8); and post-contingency (hour 9-12). Each designed period will apply a certain case of inverter failure probabilities from Table 2. For non-emergency events, all three periods follow the probability from C1, which represents a normal condition with low inverter failure probabilities. The moderate event applies C2-C5-C2 for the inverter failure probabilities of the three stages, which represent a moderate condition such as a storm. The extreme event's inverter failure probabilities follow C2-C6-C2 for the three stages, which represents an extreme condition such as a hurricane. The worst condition will result on higher inverter failure probabilities. Figure 4 illustrates the inverter failure probabilities that are calculated based on the generated scenarios. The probability patterns follow the previously mentioned cases for each event. By implementing the different events, the effectiveness of the resilience improvement of the proposed ROP algorithm is verified.

Table 4 represents the results from the implementation of the events from Figure 4. The value of SR for the non-emergency event is 96.2% which is over the pre-set value 95%, which reveals that the MG satisfies the requirement, and no extra DG is needed. For the moderate event, the value of SR drops significantly to 36% so the algorithm is adding an extra DG to the system. Note that the capacity of the extra DG is equivalent to the capacity of one of the inverter-based renewable energy sources. After implementing an extra DG to the system, the value of SR is updated to 97% which satisfies the requirement. The value of SR drops to 22.3% when testing the extreme event. Therefore, the same procedure of adding an extra DG is performed by the algorithm. However, after dispatching the extra DG, the value of SR increases to 94%, still not yielding the pre-set value. As a result, an additional DG is dispatched, and the SR value is finally improved to 99.9%.

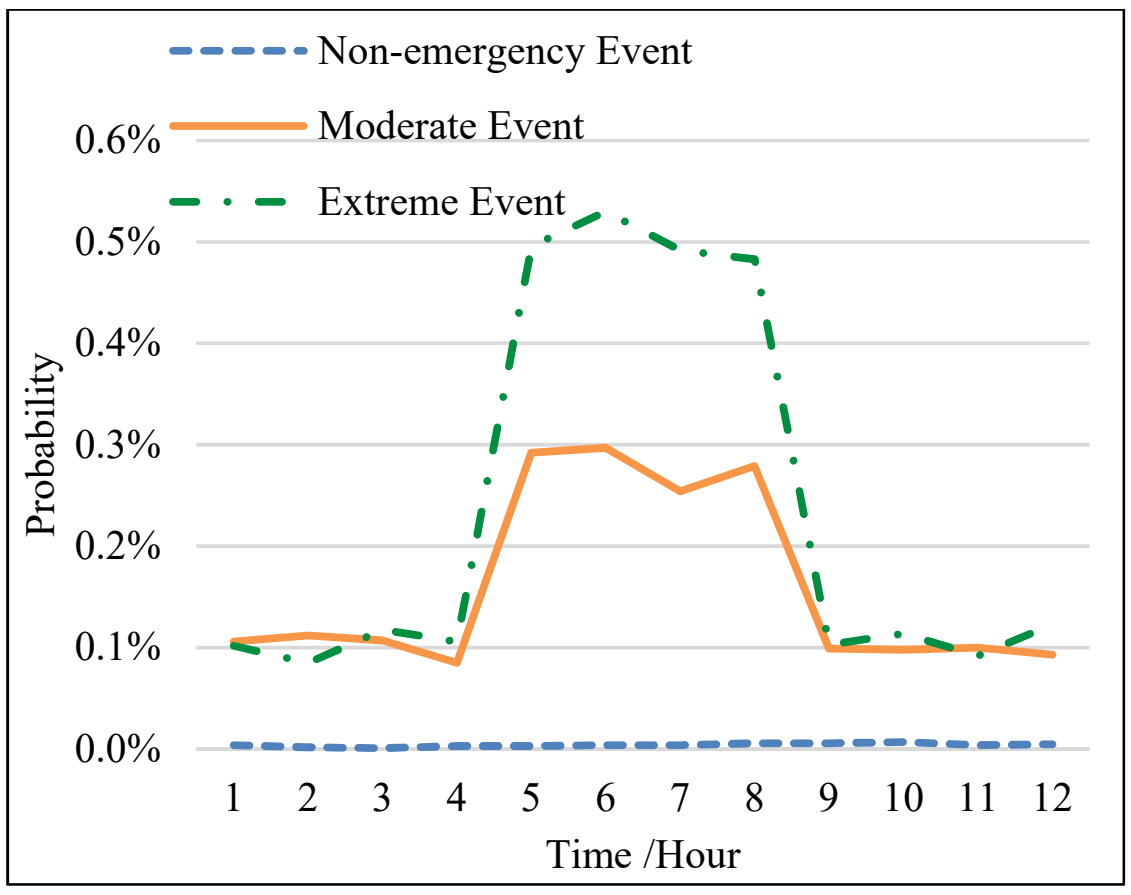

**Figure 4. Probability of at least one inverter failure under different events.**

**Table 4. Evaluation of ROP under different events.**

| SR | ROP | Add 1 DG | Add 2 DG |
|---|---|---|---|
| Non-emergency Event | 96.2% | / | / |
| Moderate Event | 36% | 97% | / |
| Extreme Event | 22.3% | 94% | 99.9% |

## 5. Conclusions

A novel resilient operational planning algorithm is proposed in this paper. The main contribution is that the proposed ROP algorithm can incorporate predicted values of inverter failure probabilities at different stages of extreme events to minimize the loss of power supply to critical loads. In addition, a microgrid survivability rate as a resilience metric is proposed to quantify the possibility that the microgrid will be able to supply enough power to its critical loads. The proposed algorithm consists of the ROP optimization step and a resilience enhancement step. In the ROP optimization step, the SR is obtained, and the resilience enhancement step further improves the SR if needed. As demonstrated in the case studies section, numerical tests validate the effectiveness of the proposed ROP optimization obtaining a higher SR value than the cost-based optimization of a traditional MEM. Also, it is demonstrated that the proposed algorithm is able to improve the SR value of an MG system by analyzing different extreme events causing different inverter failure probabilities. In summary, the proposed ROP algorithm further reinforces an MG’s ability to handle the critical load during emergency conditions.